\documentclass[aps,prb,preprint,floatfix,superscriptaddress,showpacs]{revtex4-1}

\usepackage{graphicx}
\usepackage{dcolumn}
\usepackage{amsmath,bm}
\usepackage{color}%
\usepackage{multirow}
\usepackage{array} 
\usepackage{xr}
\usepackage{xr-hyper}
\usepackage{hyperref}
\usepackage{chemformula} 
\usepackage[T1]{fontenc} 
\usepackage{braket}
\usepackage{mathtools}
\usepackage{esvect}
\usepackage{amssymb}
\usepackage{tabularx}
\usepackage{soul}
\setstcolor{red}

\newcommand*{\rom}[1]{\expandafter\@slowromancap\romannumeral #1@}
\makeatletter

\newcommand*{\addFileDependency}[1]{
  \typeout{(#1)}
  \@addtofilelist{#1}
  \IfFileExists{#1}{}{\typeout{No file #1.}}
}
\makeatother

\newcommand*{\myexternaldocument}[1]{%
    \externaldocument{#1}%
    \addFileDependency{#1.tex}%
    \addFileDependency{#1.aux}%
}

\myexternaldocument{arxiv/arxiv_ver_SI}

\begin{document}

\title{Magnetic and Crystal Symmetry Effects on Spin Hall Conductivity in Altermagnets}

\author{Dameul Jeong}
\affiliation{Department of Physics and Research Institute for Basic Sciences, Kyung Hee University, Seoul, 02447, Korea}
\author{Seoung-Hun Kang}
\email[Corresponding author. email: ]{physicsksh@khu.ac.kr}
\affiliation{Department of Physics and Research Institute for Basic Sciences, Kyung Hee University, Seoul, 02447, Korea}
\affiliation{Department of Information Display, Kyung Hee University, Seoul, 02447, Korea}
\author{Young-Kyun Kwon}
\email[Corresponding author. email: ]{ykkwon@khu.ac.kr}
\affiliation{Department of Physics and Research Institute for Basic Sciences, Kyung Hee University, Seoul, 02447, Korea}
\affiliation{Department of Information Display, Kyung Hee University, Seoul, 02447, Korea}
\date{July 2025}

\begin{abstract}
Altermagnets, which reconcile zero net magnetization with pronounced spin splitting, offer fresh opportunities for spin-based functionalities in next-generation electronic and spintronic devices. In this paper, we explore the unconventional spin Hall conductivity (USHC) in three prototypical altermagnets---RuO$_2$, CrSb, and MnTe---and elucidate how distinct magnetic and crystal symmetries modulate their spin Hall responses. RuO$_2$ exhibits only trivial USHC contributions under a tilted geometry, demonstrating that symmetry projections alone can induce apparent unconventional elements. In contrast, CrSb and MnTe manifest robust, symmetry-driven USHC without structural tilts, enabled by easy-axis orientations that reduce magnetic symmetry. Through extensive first-principles calculations, we demonstrate the complementary roles of the time-reversal-even and time-reversal-odd components in determining the overall SHC. Our findings indicate that controlling the interplay between crystal and magnetic symmetry---for instance, by epitaxial strain or doping---can provide an experimental avenue to tune USHC magnitudes and directions in altermagnets. These results pave the way for the engineering of multifunctional spintronic devices, where enhanced coherence and robust spin transport are realized in zero-net-moment materials with easily tailored spin configurations.
\end{abstract}

\maketitle

\section{Introduction}
Magnetic materials are conventionally classified as ferromagnets or antiferromagnets. Ferromagnets feature spin-split conduction bands and large net magnetic moments, while antiferromagnets exhibit no net magnetization through sublattice cancellation~\cite{alter1,alter2,alter3,alter4}. However, a new class of magnetism termed altermagnetism~\cite{alter_0} has lately attracted attention by merging the zero-net-moment property of antiferromagnets with a rotational-symmetry-enforced spin splitting within the unit cell. This singular spin ordering yields unusual transport phenomena that are different from those of conventional magnetic systems.

Altermagnetic materials hold promise for spintronics~\cite{alter_a1,alter_a2}, leveraging both the absence of stray fields (beneficial for device integration) and unconventional spin textures that can be used for advanced spin Hall effects (SHE). A paradigmatic example is RuO$_2$, which shows large SHE even with minimal spin-orbit coupling (SOC)~\cite{RuO2_2_tilt, RuO2_3}. This observation implies that exchange-driven band topology and Berry curvature make significant contributions to the spin Hall conductivity (SHC). Indeed, robust SHC in altermagnets suggests the possibility of enhanced coherence time for spins and energy-efficient spin manipulation because the net magnetic moment is absent, and thus external dipolar fields are diminished.

Beyond conventional SHC (CSHC), altermagnets can exhibit an unconventional SHC (USHC) whenever the spin current, spin polarization, and charge current lose their typical mutual orthogonality~\cite{USHC,USHC2,RuO2_2_tilt}. Although trivial USHC can arise from mere coordinate tilting, genuine (or symmetry-driven) USHC occurs when the magnetic space group forbids conventional orthogonal spin-current-charge alignments. This extra degree of freedom can enrich the device design by allowing spin-charge interconversion in geometries not realizable in ordinary ferromagnets or non-magnetic semiconductors.

In this work, we systematically address how magnetic and crystal symmetries dictate the SHC in three archetypal altermagnets: RuO$_2$, CrSb, and MnTe. We employ first-principles density functional theory (DFT) calculations incorporating SOC and Hubbard corrections, along with the Kubo formalism~\cite{SHCS_EQ1,SC_ncl_AFM2,SHC_EQ1,SHCS_EQ2,SHC_EQ2,SHCS_EQ3,USHC_ref,SHCS_EQ4}, to examine how time-reversal-even (Fermi sea) and time-reversal-odd (Fermi surface) terms reshape the SHC in varying symmetry settings. Our main findings are as follows. First, RuO$_2$---although previously highlighted as an altermagnet---yields only \emph{trivial} USHC when the crystal is oriented to break orthogonality. Second, CrSb and MnTe display significant intrinsic USHC that emerges from genuine magnetic symmetry breaking rather than mere structural tilts. Third, these results suggest experimentally testable scenarios in which doping, strain engineering, or finite-temperature disorder can be exploited to tune the interplay of crystal and magnetic symmetries. Finally, we outline a plausible route to implement these findings in device architectures, such as spin-orbit torque geometries, where the absence of net moment in altermagnets can reduce unwanted cross-interactions.

\section{Computational Methodology}
\label{sec:comput}
\subsection{DFT Calculations}
Our theoretical framework is based on DFT~\cite{hohenberg1964inhomogeneous,kohn1965self} implemented in Quantum Espresso~\cite{QE-2009,QE-2017}. Projector Augmented Wave pseudopotentials~\cite{blochl1994projector} were adopted, with an energy cutoff of 60~Ry for wavefunctions and the PBE-GGA exchange-correlation functional~\cite{perdew1996generalized}. The Brillouin-zone integrals were performed using Monkhorst-Pack grids of $16\times16\times16$, $15\times15\times9$, and $15\times15\times7$ $k$-points for RuO$_2$, CrSb, and MnTe, respectively. SOC was included self-consistently, and an additional Hubbard $U$ term was introduced for Ru $4d$ (2~eV) and Mn $3d$ (4~eV) orbitals to capture on-site correlations~\cite{HubbardU,HubbardU2,alter_0,RuO2_2_tilt,li2019intrinsic,chen2019topological,lai2021defect}.

Following ground-state computations, we used maximally localized Wannier functions~\cite{marzari1997maximally,mostofi2014updated} to construct tight-binding Hamiltonians for precise interpolation of band structures and spin transport quantities. The SHC was calculated within WannierBerri~\cite{tsirkin2021high} on a dense $100\times100\times100$ $k$-mesh and a Fermi level broadening $\Gamma$ $\approx$ 50~meV.

\subsection{Spin Hall Conductivity Formalism}
SHC is a key parameter for spintronics systems, which facilitates spin-to-spin and spin-to-charge interconversion. Theoretical estimations of SHC rely on the Kubo formalism, as refined in previous work, to explicitly incorporate material symmetries~\cite{SHCS_EQ1, SC_ncl_AFM2, SHC_EQ1, SHCS_EQ2, SHC_EQ2, SHCS_EQ3, USHC_ref, SHCS_EQ4}. Within this framework, the SHC emerges from two primary contributions, the Fermi surface and sea terms, each sensitive to different material symmetries. The Fermi surface term applies predominantly to magnetic systems and focuses on states near the Fermi level. Formally, it is defined by
\begin{equation}
\mathcal{\sigma}^{odd,k}_{ij} = \frac{-e\hbar}{2\pi}{\Gamma^2}\int_{BZ}\frac{d^3k}{(2\pi)^3}\sum_{n,m}\frac{\text{Re}(\braket{n\mathbf{k}|\frac{1}{2}\{\hat{s}_k,\hat{v}_i\}|m\mathbf{k}}\braket{m\mathbf{k}|\hat{v}_j|n\mathbf{k}})}{[(E_{F}-\epsilon_{n\mathbf{k}})^2+\Gamma^2][(E_{F}-\epsilon_{m\mathbf{k}})^2+\Gamma^2]},
\label{Surface}
\end{equation}
where $\Gamma$ represents the band broadening constant. This term is highly sensitive to the magnetization direction, exhibiting odd symmetry under magnetization reversal, which highlights its relevance in systems where the SHC depends on magnetic symmetry. 
In contrast, the Fermi sea contribution is relevant primarily in non-magnetic systems, as it is associated with states deeply embedded in the Fermi sea. The influence of this term on SHC is expressed as
\begin{equation}
\mathcal{\sigma}^{even,k}_{ij} = {-e\hbar}\int_{BZ}\frac{d^3k}{(2\pi)^3}\sum_{n,m}(f_{n\mathbf{k}}-f_{m\mathbf{k}})\frac{\text{Im}(\braket{n\mathbf{k}|\frac{1}{2}\{\hat{s}_k,\hat{v}_i\}|m\mathbf{k}}\braket{m\mathbf{k}|\hat{v}_j|n\mathbf{k}})}{(\epsilon_{n\mathbf{k}}-\epsilon_{m\mathbf{k}})^2},
\label{Sea}
\end{equation}
It maintains even symmetry with respect to magnetization reversal. This symmetry enables the Fermi sea term to characterize SHC in non-magnetic crystals via symmetry properties independent of magnetization.

Through the combined framework of the Fermi surface and sea terms, SHC can be systematically analyzed across a broad range of materials. The Fermi surface term aligns with magnetic symmetry, while the Fermi sea term predominantly reflects non-magnetic crystal symmetries, allowing for the tailored manipulation of SHC based on intrinsic material symmetries~\cite{SHCS_EQ1, SHCS_EQ2, SHCS_EQ3, SHCS_EQ4}.

The SHC tensor, represented as a third-order tensor, comprises 27 components, with indices \( i \), \( j \), and \( k \) independently aligned along the \( x \), \( y \) or \( z \) axes. CSHC components arise only when \( i \), \( j \), and \( k \) are mutually orthogonal; other configurations are defined as USHC terms.

To simplify analysis, we introduce the reduced quantity \( S^k_{ij} = v_i \sigma_k v_j \)~\cite{USHC}, where, $v_{i,j}$ are defined as $\frac{1}{\hbar}(\frac{\partial{H}}{\partial{k_{i,j}}})$, and \( \sigma_{k} \) represents the Pauli spin matrices. Here, \( S^k_{ij} \) captures the essence of the SHC tensor \( \sigma^{k}_{ij} \) while allowing direct symmetry evaluations. A symmetry operator \( \mathcal{O} \) allows a component for \( \mathcal{O} S^k_{ij} = S^k_{ij} \), while it forbids the component for \( \mathcal{O} S^k_{ij} = -S^k_{ij} \). If any symmetry operator on some physical quantity in the Fermion system produces a negative value, that physical quantity is not allowed. This analysis shows that while the even component \( \sigma^{\text{even},k}_{ij} \) depends solely on crystal symmetry, the odd component \( \sigma^{\text{odd},k}_{ij} \) also additionally necessitates magnetic symmetry considerations due to its time-reversal asymmetry.

\section{Results and Discussion}
\subsection{Structural and Spin Orientations in Altermagnets}

\begin{figure}[t!]
\includegraphics[width=1\columnwidth]{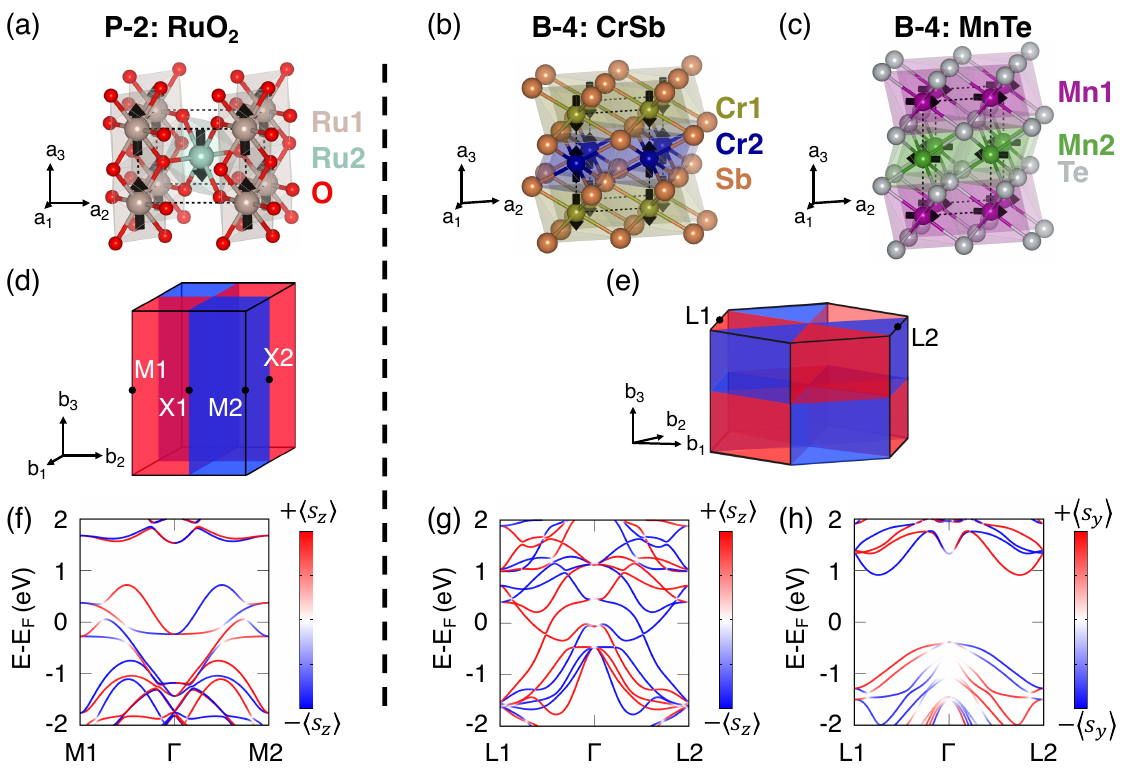}
\caption{Crystal structures for a) RuO$_2$, b) CrSb, and c) MnTe, with lattice vectors \(\textbf{a}_1\), \(\textbf{a}_2\), and \(\textbf{a}_3\) (\(\textbf{a}_1\) along the $x$-axis). Brillouin zones of d) P-2 and e) B-4 symmetries with spin-momentum locking, where reciprocal lattice vectors \(\textbf{b}_1\), \(\textbf{b}_2\), and \(\textbf{b}_3\) (\(\textbf{b}_1\) along $k_x$ in (d) and \(\textbf{b}_2\) along $k_y$ in (e)). Red and blue denote opposite spin-momentum locking directions. Spin-orbit coupling (SOC)-resolved band structures for f) RuO$_2$, g) CrSb, and h) MnTe show spin projections: red/blue lines indicate positive/negative spin projections along $z$ (RuO$_2$, CrSb) or $y$ (MnTe).
\label{FIG1}}
\end{figure}
RuO$_2$, CrSb, and MnTe are recognized altermagnets, each exhibiting unique magnetic properties. Although the magnetic nature of RuO$_2$ remains debated~\cite{RuO2_alter_O1,RuO2_alter_O2,RuO2_alter_O3,RuO2_alter_O4,RuO2_alter_X1,RuO2_alter_X2,RuO2_alter_X3,RuO2_alter_X4,RuO2_alter_X5,song2025spin}, we assumed it to be an altermagnet for comparison with the SHC results from previous studies~\cite{RuO2_2_tilt}, as detailed in the Supplementary Note S2. Figure~\ref{FIG1}a-c depict the crystal structures of RuO$_2$, CrSb, and MnTe, along with their respective easy axes identified in previous works~\cite{RuO2_easy_c1,RuO2_easy_c2,CrSb_easy_c1,CrSb_easy_c2,CrSb,MnTe_exp,MnTe_easy_y1}. According to the classification by {\v{S}}mejkal \emph{et al.}\cite{alter_0}, RuO$_2$ is a planar altermagnet (P-2), while CrSb and MnTe fall into the bulk category (B-4). Figure~\ref{FIG1}d-e schematically illustrate spin-momentum locking within the first Brillouin zone for the category. RuO$_2$, CrSb, and MnTe demonstrate robust spin splitting despite SOC-induced perturbations~\cite{alter_SOC}. As shown in Figure~\ref{FIG1}f-h, each compound exhibits clear spin polarization in its band structure, with RuO$_2$ and CrSb spins aligned along $\hat{z}$, and MnTe along $\hat{y}$ ($[01\bar{1}0]$) in our coordinate system. A detailed comparison of band structures with and without SOC is provided in the Supplementary Note S1.

\subsection{Trivial USHC in Tilted RuO$_2$}
\begin{figure}[t!]
\includegraphics[width=1\columnwidth]{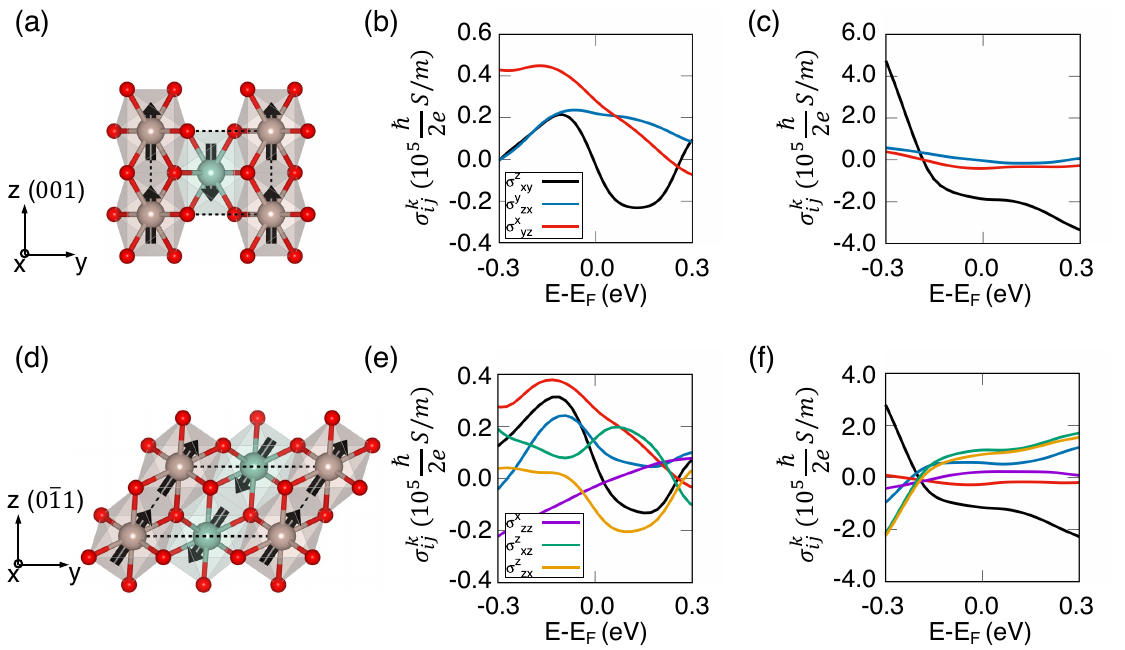}
\caption{Fermi sea (time-reversal even) and surface (time-reversal odd) contributions to the spin Hall conductivity (SHC) of RuO$_2$ along the (001) and ($0\bar{1}1$) orientations. Side views of RuO$_2$ with magnetic moments aligned along [001] for the a) (001) and d) ($0\bar{1}1$) orientations, respectively. b) The sea and c) surface contributions to the SHC for the (001)-oriented RuO$_2$, where symmetry only allows conventional SHC (CSHC) elements. Rotation symmetry allows six CSHC elements, reducible to three independent components. e) The sea and f) surface contributions to SHC for the ($0\bar{1}1$)-oriented RuO$_2$. The tilted orientation and spin alignment break symmetry, introducing unconventional SHC (USHC) elements. However, these contributions are trivial due to identical results from both sea and surface terms.
\label{FIG2}}
\end{figure}


In RuO$_2$, SHC arises from distinct contributions of the Fermi sea and surface, constrained by the 4/mmm Laue symmetry in the crystal. Figure~\ref{FIG2}a shows spin aligned with $z$-direction((001) axis), where only CSHC elements contribute from the Fermi sea. Figure~\ref{FIG2}b,c highlight these CSHC terms corresponding to spin polarizations in $x$, $y$, and $z$. The non-symmorphic symmetry, involving glide-reflection symmetry as a combination of mirror symmetry $(\mathcal{M}_{x,y})$ with half translations $(\tau_{y+1/2,x+1/2} + \tau_{z+1/2})$ (see the Supplementary Note S3), imposes constraints on the components of SHC to be only CSHC.

In contrast, a ($0\bar{1}1$)-oriented RuO$_2$ introduces a structural tilt, as shown in Figure~\ref{FIG2}d, allowing USHC components. This tilt alters both structural and spin directions, allowing trivial USHC contributions from both the Fermi sea and the surface, as shown in Figure~\ref{FIG2}e and f. Trivial USHC refers to non-intrinsic contributions arising from geometric or structural effects rather than symmetry-driven mechanisms. The trivial nature of USHC is evidenced by the identical matrix elements derived from both Fermi sea and surface contributions, regardless of the tilting angle. All SHC tensor elements for tilted structures can be determined analytically with a tilting angle, removing the need for direct DFT calculation.

\subsection{Symmetry-Driven Intrinsic USHC in CrSb and MnTe}
\begin{figure}[t!]
\includegraphics[width=1.0\columnwidth]{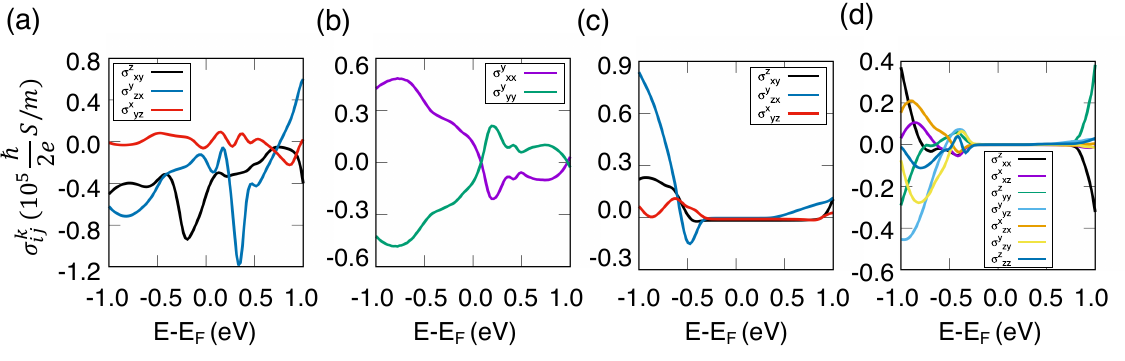}
\caption{Magnetic symmetry-Driven USHC Variations in CrSb and MnTe. a) Fermi sea and b) Fermi surface contributions to the SHC in CrSb; c) Fermi sea and d) Fermi surface contributions to the SHC in MnTe. For both CrSb and MnTe, the Fermi sea contribution permits only CSHC elements. However, when magnetic symmetry is applied to the Fermi surface contribution, CSHC components are disallowed. Although CrSb and MnTe share the same crystal structure, their differing magnetic symmetries, due to distinct, easy axes, lead to variations in the permissible USHC elements. Moreover, MnTe lacks magnetic rotational symmetry, resulting in all USHC elements being independent of each other.
\label{FIG3}}
\end{figure}


In contrast, CrSb and MnTe display distinct USHC characteristics that separate them from RuO$_2$. As shown in Figure~\ref{FIG3}a,c , both materials belong to the $6/mmm$ Laue group, which confines Fermi sea (even) contributions to CSHC components, resembling the behavior of non-magnetic systems with the same symmetry group~\cite{USHC}. These contributions are primarily governed by crystal symmetry. In contrast, as shown in Figure~\ref{FIG3}b,d , the Fermi surface (odd) contributions in CrSb and MnTe diverge from the constraints of CSHC, permitting only specific USHC matrix elements.

In CrSb, uniaxial magnetization along $[0001]$ breaks certain mirror planes that would otherwise preserve orthogonality. Consequently, spin-orbit coupling lifts degeneracies in the nodal plane $k_z=0$, giving rise to strong in-plane spin textures. The components $\sigma^{\text{odd},y}_{xx}$, $\sigma^{\text{odd},x}_{yx}$, and $\sigma^{\text{odd},x}_{xy}$ are equivalent due to constraints imposed by rotational symmetry $C_{3z}$. Meanwhile, in MnTe, the spins lie in the $[01\bar{1}0]$ direction, further reducing the effective magnetic symmetry. This yields an even broader set of symmetry-allowed USHC elements, as illustrated in Figure~\ref{FIG3}d, each of which can have distinct magnitudes due to the lack of higher-order rotational equivalences.

\subsection{Band Structures and Spin Textures}
\begin{figure}[htb!]
\includegraphics[width=1\columnwidth]{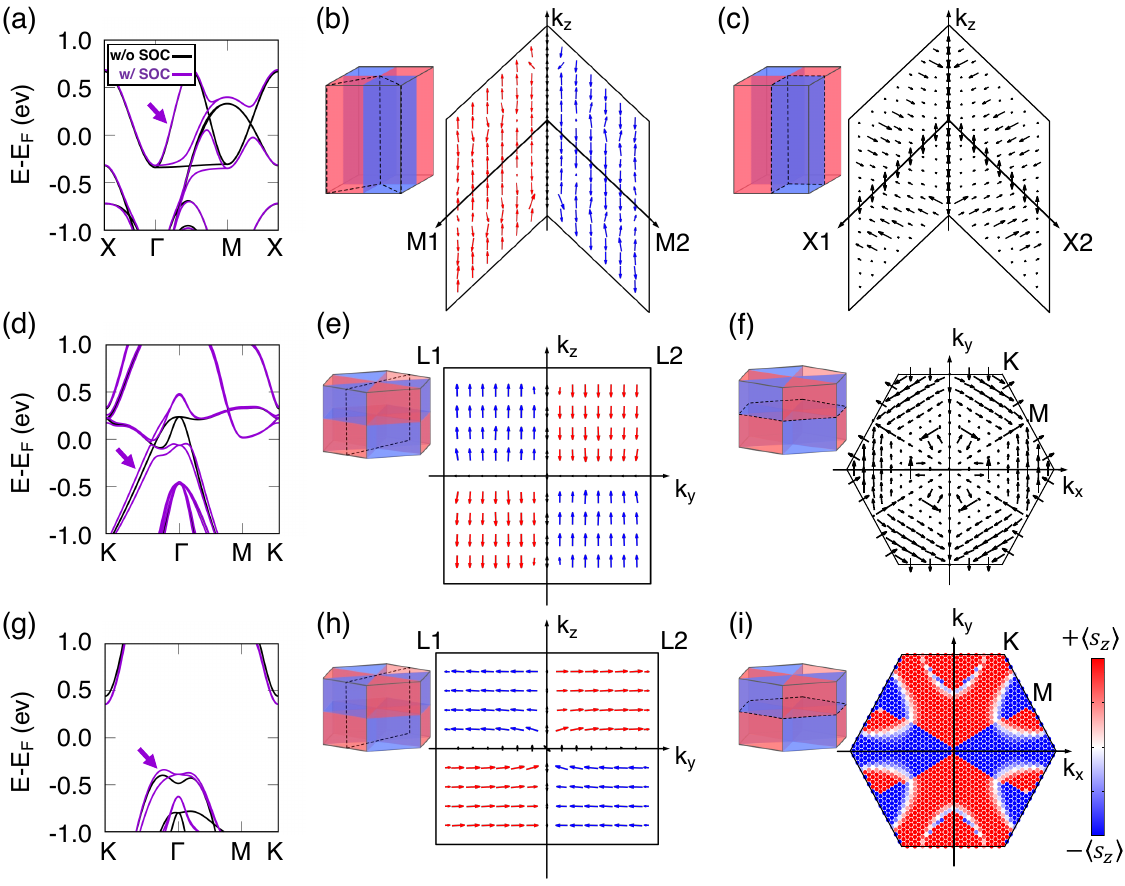}
\caption{Band structure on the ($k_x, k_y, k_z=0$) plane and spin texture in the Brillouin zone for RuO$_2$, CrSb, and MnTe. The spin texture corresponds to the band indicated by the arrow. a) RuO$_2$ exhibits spin degeneracy along the $X$-$\Gamma$ and $M$-$X$ nodal lines, with altermagnetic spin splitting along $\Gamma$-$M$. b) Spin texture on the plane formed by $\Gamma$-$M1$, $\Gamma$-$M2$, and $k_z$ for bands highlighted in (a). Red and blue indicate opposite spin-momentum locking directions, and a spin-degenerate nodal line along $k_z$ is shown in the black line. c) Spin texture on the plane is defined by $\Gamma$-$X1$, $\Gamma$-$X2$, and $k_z$. d) For CrSb, spin splits along all paths in the ($k_x, k_y, k_z=0$) plane with SOC. e) Spin texture on the $k_y$-$k_z$ plane for bands highlighted in (d). f) Spin texture in CrSb shows only in-plane spin. g) In MnTe, spin splits along all paths within the ($k_x, k_y, k_z=0$) plane. h) Spin texture on the $k_y$-$k_z$ plane for bands highlighted in (g). i) The nodal plane in MnTe shows an out-of-plane spin component; red denotes spin in the $+z$ direction, and blue in the $-z$ direction.
\label{FIG4}}
\end{figure}

Figure~\ref{FIG4} explores the band structure and spin texture in RuO$_2$, CrSb, and MnTe, emphasizing the role of symmetry and SOC. Figure~\ref{FIG4}a shows the RuO$_2$ band structure along the $(k_x, k_y, k_z=0)$ plane, highlighting spin splitting along $\Gamma$-$M$ with and without SOC and symmetry-enforced spin degeneracy at nodal lines. Figure~\ref{FIG4}b,c (projected from the band indicated by the arrow in Figure~\ref{FIG4}a) illustrate the spin textures on planes defined by $k_z$ and directions $M1$, $M2$ ($[\bar{1}10]$, $[110]$) and $X1$, $X2$ ($[010]$, $[100]$), respectively, revealing spin-momentum locking in the relativistic regime and spin degeneracy along high-symmetry axes such as $\Gamma$-$X1$, $X2$, and $Z$.

In contrast, CrSb exhibits bulk spin-momentum locking, as shown in Figure~\ref{FIG4}e, derived from the band marked in Figure~\ref{FIG4}d. Under non-magnetic conditions, CrSb shares mirror symmetry planes with RuO$_2$, including $(k_x=0, k_y, k_z)$, $(k_x, k_y=0, k_z)$, and $(k_x, k_y, k_z=0)$. The glide-reflection symmetry ($\mathcal{M}_{y}(\tau_{x}+\tau_{z})$) preserves the $(k_x, k_y=0, k_z)$ plane, as detailed in Supplementary Note S3. However, magnetic symmetry breaks mirror planes such as $(k_x=0, k_y, k_z)$ and $(k_x, k_y, k_z=0)$, leading to SOC-driven spin splitting in the $k_z=0$ plane, as reflected in Figure~\ref{FIG4}d and the in-plane spin texture in Figure~\ref{FIG4}f.

MnTe shares the same crystal symmetry as CrSb, but exhibits distinct magnetic symmetry due to spin alignment along the [01$\bar{1}$0] direction. Figure~\ref{FIG4}h, based on the band marked in Figure~\ref{FIG4}g, shows spins predominantly aligned along the $y$-axis. With SOC, spin degeneracy is lifted in the $(k_x, k_y, k_z=0)$ plane, causing spin alignment along $k_z$, as seen in Figure~\ref{FIG4}i. This change in spin direction reduces the symmetry of the system. Moreover, unlike RuO$_2$ or CrSb, MnTe lacks the glide-reflection symmetry ($\mathcal{M}_{y}(\tau_{x+1/2}+\tau_{z+1/2})$) but retains the mirror symmetry ($\mathcal{M}_z$), preserving $\sigma_z$ under its operation. This reduced symmetry, resulting from the [01$\bar{1}$0] spin orientation, enables MnTe to exhibit a broader range of non-equivalent USHC tensor elements compared to CrSb.

Analytical symmetry operations confirm the unique SHC components of MnTe. There are two non-trivial symmetry operations in this system to evaluate SHC tensor elements. The symmetry operation combined with time-reversal and a 180$^\circ$ rotation about the $x$-axis $\mathcal{T}C_{2x}$ and mirror operation $\mathcal{M}_z$. Among them, $\mathcal{T}C_{2x}$ forbids all CSHC components while allowing diverse USHC elements, including $\sigma^{\text{odd},z}_{xx}$, $\sigma^{\text{odd},x}_{xz}$, $\sigma^{\text{odd},z}_{yy}$, $\sigma^{\text{odd},x}_{zx}$, $\sigma^{\text{odd},y}_{zy}$, and $\sigma^{\text{odd},z}_{zz}$. The spin texture in Figure~\ref{FIG4}h further illustrates spin alignment along the $y$-axis, with a $k_z$ component appearing upon inclusion of SOC. This orientation reduces symmetry compared to RuO$_2$ and CrSb, leading to a greater diversity of USHC tensor elements. Detailed transformations and equations are provided in Supplementary Note S4.


\subsection{Experimental and Device Design Outlook}
Although our work is primarily theoretical, there are several tangible routes to verify and utilize these predicted USHC phenomena in altermagnets:

\textbf{Epitaxial Growth and Strain Control:}  
 Strain in RuO$_2$ can be effectively controlled by epitaxial growth on TiO$_2$~\cite{jeong2025anisotropic,jeong2025metallicity}. Similarly, CrSb and MnTe have also been successfully grown on widely utilized substrates such as GaAs and sapphire~\cite{CrSb, MnTe_exp,chilcote2024stoichiometry,aota2025epitaxial}, where substrate-induced strain plays a critical role in tuning their magnetic anisotropy. Our results suggest that small shifts in the easy-axis orientation can move the system between distinct USHC regimes. By measuring spin Hall signals via lock-in detection of spin-torque ferromagnetic resonance (ST-FMR) or second-harmonic Hall, experimentalists can map the evolution of USHC components under systematically varying strain.

\textbf{Chemical Doping or Alloying:}  
Introducing dopants or forming solid solutions (e.g., (Cr, M) Sb with M=Mn, Fe, Co, Ni and V or Mn (Te, X) with X = Se, S) can shift the Fermi level and modulate the band structure near critical nodal planes. Because the Fermi surface term dominates the time-reversal-odd contribution, doping can drastically alter the magnitude and sign of USHC. Recent experiments on MnTe thin films show that stoichiometric tuning, such as excess Mn incorporation during epitaxial growth, can drive the system towards weak ferromagnetism and metallicity, offering a direct handle on altermagnetic transport signatures~\cite{chilcote2024stoichiometry}. Such doping studies can also clarify whether the predicted spin-split bands indeed drive the unconventional spin Hall response.

\textbf{Spintronic Device Architectures:}  
Altermagnets offer zero net moment and strong spin-split bands—ideal for spin-orbit torque (SOT) devices. Recent studies of USHC in low-symmetry systems (e.g., IrO$_2$)~\cite{yang2025coexistence} using ST-FMR have shown that unconventional spin current components can be generated and detected efficiently Analogous bilayer devices combining heavy metals with CrSb or MnTe could enable current-driven spin injection and re-emission along unconventional directions, enabling multi-functional operations controllable by strain or doping.


\section{Conclusions}
In this study, we systematically investigated the unconventional spin Hall conductivity (USHC) in representative altermagnetic materials, such as RuO$_2$, CrSb, and MnTe, highlighting the critical roles played by their distinct magnetic and crystal symmetries. RuO$_2$ was shown to exhibit only trivial USHC contributions induced by structural tilting, without genuine intrinsic unconventional features. In contrast, CrSb and MnTe demonstrated significant intrinsic USHC arising explicitly from their reduced magnetic symmetry. Our results underline how different magnetic easy-axis orientations, combined with crystal symmetry breaking, strongly influence both time-reversal-even (Fermi-sea) and time-reversal-odd (Fermi-surface) SHC contributions.

These findings open new opportunities for the engineering of advanced spintronic devices using altermagnets. Specifically, we propose experimental pathways, such as epitaxial strain control, doping, and alloying, to manipulate and optimize the magnitude and directionality of USHC. Such tunability of spin currents in materials with zero net magnetization provides a significant advantage, reducing undesirable stray fields and enhancing spin coherence. Ultimately, our work establishes a clear theoretical foundation for the experimental realization of novel altermagnet-based spintronic applications, promoting energy-efficient device designs and versatile spin-charge interconversion mechanisms.

\section{Acknowledgements}
This research was supported by the Korean government (MSIT) through the National Research Foundation of Korea (2022R1A2C1005505, RS-2024-00416976). Our computational work was partially done using the resources of the KISTI Supercomputing Center (KSC-2024-CRE-0540).



\bibliographystyle{rsc}
\bibliography{arxiv/bib} 

\end{document}